\documentclass[aps,prl,superscriptaddress]{revtex4}

\usepackage{amsmath}
\usepackage{graphicx}

\usepackage{amsmath}
\usepackage{graphicx}

\begin{document}

\title{Supplementary material to:
Finite-size effects in the interfacial stiffness of
rough elastic contacts}

\author{Lars Pastewka}
\affiliation{Dept. of Physics and Astronomy, Johns Hopkins University, Baltimore, MD 21218, USA}
\affiliation{MikroTribologie Centrum $\mu$TC, Fraunhofer-Institut f\"ur Werkstoffmechanik IWM, Freiburg, 79108 Germany}
\author{Nikolay Prodanov}
\affiliation{J\"ulich Supercomputing Center, Institute for Advanced Simulation, FZ J\"ulich, 52425 J\"ulich, Germany}
\affiliation{Dept. of Materials Science and Engineering, Universit\"at des Saarlandes, 66123 Saarbr\"ucken, Germany}
\author{Boris Lorenz}
\affiliation{Peter Gr\"unberg Institut-1, FZ-J\"ulich, 52425 J\"ulich, Germany}
\author{Martin H. M\"user}
\affiliation{J\"ulich Supercomputing Center, Institute for Advanced Simulation, FZ J\"ulich, 52425 J\"ulich, Germany}
\affiliation{Dept. of Materials Science and Engineering, Universit\"at des Saarlandes, 66123 Saarbr\"ucken, Germany}
\author{Mark O. Robbins}
\affiliation{Dept. of Physics and Astronomy, Johns Hopkins University, Baltimore, MD 21218, USA}
\author{Bo N. J. Persson}
\affiliation{Peter Gr\"unberg Institut-1, FZ-J\"ulich, 52425 J\"ulich, Germany}

\begin{abstract}
In this supplementary materials section, we provide
(i) additional information on the numerical simulations of the main work,
(ii) the derivation of all prefactors in the analytical theory, and
(iii) unpublished experiments of the contact stiffness of a polymer pressed
against a rough substrate.
\end{abstract}

\maketitle

\pagestyle{empty}


\section{Introduction}

In this supplementary paper we present some details which could no be included in the PRL because of space limitations.
We first show the surface roughness power spectra used in the simulations. Next we derive the prefactor in the power law relation
between the contact stiffness and the load. Finally we describe a new measurement of the contact stiffness where a silicon rubber block is pressed against an
asphalt road surface.

\section{Numerical details}

\begin{figure}
\includegraphics[width=0.5\columnwidth]{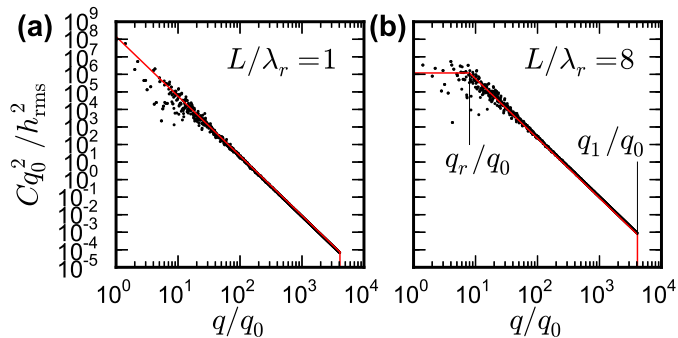}
\caption{\label{fig:power_spectrum}
(Color online) Power spectra for two surfaces without (a) and with (b) a roll-off at large wavelength as generated by a Fourier filtering algorithm. The solid lines show the prescribed power spectrum $C(q)$ and the dots the actual realization. Panel (b) indicates the wavevectors of the long-wavelength roll-off $q_r=\pi/\lambda_r$ and the short-wavelength cutoff $q_1=\pi/\lambda_1$. For $q<q_0=\pi/L$ where $L$ is the linear system size the surfaces have zero power. The noise at low $q$ is due to the fact that order $q^2$ Fourier components contribute to the power-spectrum of a realization of a surface.
}
\end{figure}

Fig.~\ref{fig:power_spectrum} shows a surface roughness power spectrum as used
in the simulations.
The solid lines indicate the mean values for the spectrum, while the dots
reflect one particular realization.
Fluctuations of the height $h({\bf r})$ in real space are not only the consequence of
variations in the absolute value of their complex
Fourier transforms $\tilde{h}({\bf q})$
but also due to the random phases.

\section{Derivation of prefactors}

\begin{figure}[thbp]
\includegraphics[width=0.35\columnwidth]{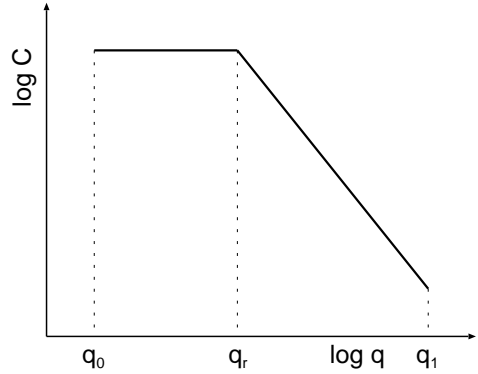}
\caption{\label{roll}
The surface roughness power spectra as a function of the wavevector (log-log scale) for a self-affine fractal
surface with a roll-off.
}
\end{figure}

Consider a randomly rough surface with a roll-off as indicated in Fig. \ref{roll}. The power spectrum
$$C(q) = C_0 \ \ \ \ {\rm for} \ \ \ \ q_0 < q < q_r$$
$$C(q)=C_0 \left ({q\over q_r}\right )^{-2(1+H)}  \ \ \ \ {\rm for} \ \ \ \ q_r < q < q_1$$
where $q_0=\pi /L$, where $L$ is the linear size of the studied system. We also write $q_r = \pi /\lambda_r$ where $\lambda_r$
is the roll-off wavelength. The surface mean square roughness amplitude
$$(h^o_{\rm rms})^2 = \int d^2 q \ C(q) = 2\pi C_0 \left [\int_{q_0}^{q_r}
dq \ q+ \int_{q_r}^{q_1} dq \ q \left ({q\over q_r}\right )^{-2(1+H)}\right ] \approx
{\pi q_r^2 \over Hs} C_0 $$
where
$$1/s=1+H\left [1-(q_0/q_r)^2\right ]$$
Thus
$$C_0 = {Hs\over \pi q_r^2} (h^o_{\rm rms})^2$$

We first calculate the elastic energy stored in the deformation field associated with the Hertz mesoscale asperity contact region.
The mesoscale asperity has the radius of curvature $R$ and the radius of the (apparent) contact region between the mesoscale asperity and the
flat countersurface is denoted by $r_0$.
The mean summit asperity curvature is given by~\cite{Nayak}
$\bar \kappa = \beta \surd 2 \kappa_0$
where $\kappa_0$ is the root-mean-square curvature of the surface
$$\kappa^2_0 = {1\over 2} \int d^2q \ q^4 C(q) = \pi \int_{q_0}^{q_1} dq \ q^5 C(q)$$
When roughness occurs on many length scales so that $q_1/q_0 >> 1$ Nayak has shown that $\beta = (8/3\pi)^{1/2}$.
Including only roughness components with wavevector $q < \pi /r_0$ gives the mean summit curvature $1/R$ of the mesoscale asperity:
$${1\over R^2} = 2\pi \beta^2  \int_{q_0}^{\pi /r_0} dq \ q^5 C(q)$$
We define $\bar R = q_rR$, $\bar h^o_{\rm rms} = q_r h^o_{\rm rms}$ and $\bar r_0 = q_r r_0$ and assume $\pi /r_0 > q_r$. Thus
$${1\over \bar R^2} = 2\pi C_0 q_r^{-2}\left [\int_{q_0}^{q_r} dq \ q^5 +\int_{q_r}^{\pi/r_0} dq \ q^5 \left ({q\over q_r}\right )^{-2(1+H)}\right ]
\approx
2 \pi \beta^2 q_r^4 C_0 \int_{1}^{\pi /\bar r_0} dx \ x^{3-2H} \approx {Hs \beta^2\over 2-H} (\bar h^o_{\rm rms})^2 \left ({\pi \over \bar r_0} \right )^{4-2H}$$
and
$${1\over \bar R} = \left ({Hs\beta^2 \over 2-H}\right )^{1/2} \bar h^o_{\rm rms} \pi^{2-H} \bar r_0^{H-2}=\alpha^{-1} \bar r_0^{H-2}$$
where
$$\alpha = \left ({Hs\beta^2 \over 2-H}\right )^{-1/2} \left ( \bar h^o_{\rm rms} \right )^{-1} \pi^{H-2}$$
Thus
$$\bar R = \alpha \bar r_0^{2-H}$$

From Hertz theory
$$r_0^3 = {3FR \over 4E^*}$$
Defining
$$\bar F = {Fq_r^2 \over E^*}$$
gives
$$\bar r_0^3 = {3\over 4} \bar F \bar R = {3 \alpha \over 4} \bar F \bar r_0^{2-H}$$
$$\bar r_0 = \left ( {3 \alpha \over 4}\right )^{1/(1+H)} \bar F^{1/(1+H)}$$
Thus
$$\bar R = \alpha \left ({3\alpha \over 4}\right )^{(2-H)/(1+H)} \bar F^{(2-H)/(1+H)}$$
or
$$\bar R = \left ({3\over 4}\right )^{(2-H)/(1+H)} \alpha^{3/(1+H)} \bar F^{(2-H)/(1+H)}$$

The elastic energy stored in the Hertz mesoscale deformation field:
$$U_{\rm el}^{(0)} = {2\over 5} F \delta = {2\over 5} E^* q_r^{-3} \bar F \bar \delta $$
where $\bar \delta = q_r \delta$. We define
$$\bar U_{\rm el}^{(0)} = U_{\rm el}^{(0)}q_r^3/E^*$$
so that
$$\bar U_{\rm el}^{(0)} = {2\over 5} \bar F \bar \delta $$
Next
$$\bar \delta = q_r \delta = q_r \left ({9 F^2 \over 16 RE^{*2}}\right )^{1/3} =\left ( {9\bar F^2\over 16 \bar R}\right )^{1/3}
=\left ( {9\over 16}\right )^{1/3} \left ({3\over 4}\right )^{(H-2)/3(1+H)} \alpha^{-1/(1+H)} \bar F^{H/(1+H)}$$
Thus
$$\bar U_{\rm el}^{(0)} =
{2\over 5} \left ( {9\over 16}\right )^{1/3} \left ({3\over 4}\right )^{(H-2)/3(1+H)} \alpha^{-1/(1+H)} \bar F^{(1+2H)/(1+H)} = {2\over 5} \left ({3\over 4}\right )^{H/(1+H)} \alpha^{-1/(1+H)} \bar F^{(1+2H)/(1+H)}$$

Next we calculate the elastic deformation energy stored in the vicinity of the microasperity contact
regions in the Hertz mesoscale contact region~\cite{Persson,Camp}:
$$U_{\rm el}^{(1)} = u_1 A p_1 = u_1 F$$
or
$$\bar U_{\rm el}^{(1)} = q_r^3 U_{\rm el}^{(1)}/E^* = \bar u_1 \bar F$$
where $\bar u_1 = q_r u_1$. We have $\bar u_1 = \gamma \bar h_{\rm rms}$ where $\gamma \approx 0.4$ and
$$(h_{\rm rms})^2 = \int d^2 q \ C(q) = 2\pi C_0 \int_{\pi/r_0}^{q_1} dq \ q \left ({q\over q_r}\right )^{-2(1+H)} \approx
2\pi C_0 q_r^2 q_r^{2H} (\pi /r_0)^{-2H} (2 H)^{-1}=  (h^o_{\rm rms})^2 s \pi^{-2H} r_0^{2H} q_r^{2H}$$
or
$$\bar h_{\rm rms} = \bar h^o_{\rm rms} s^{1/2} \pi^{-H} \bar r_0^{H} = \bar h^o_{\rm rms} s^{1/2}  \pi^{-H}
\left ({3\alpha \over 4}\right )^{H/(1+H)} \bar F^{H/(1+H)}$$
Using the definition for $\alpha$ to eliminate $ \bar h^o_{\rm rms}$ we can write
$$\bar h_{\rm rms} =
\left ({3 \over 4}\right )^{H/(1+H)} \left ({2-H\over H\beta^2 }\right )^{1/2} \pi^{-2} \alpha^{-1/(1+H)}  \bar F^{H/(1+H)}$$
Thus
$$\bar U_{\rm el}^{(1)} = \bar u_1 \bar F = \gamma \bar h_{\rm rms} \bar F =
\gamma \left ({3\over 4}\right )^{H/(1+H)} \left ( {2-H\over H\beta^2}\right )^{1/2} \pi^{-2}\alpha^{-1/(1+H)} \bar F^{(1+2H)/(1+H)}$$
The total elastic energy
$$\bar U_{\rm el} = \bar U^{(0)}_{\rm el}+\bar U^{(1)}_{\rm el} =
\left ({3\over 4}\right )^{H/(1+H)} \alpha^{-1/(1+H)} \bar F^{(1+2H)/(1+H)} \left [{2\over 5} +
 \left ( {2-H\over H}\right )^{1/2} {\gamma \over \beta \pi^2}\right ]$$

The total stiffness $k=K A_0$ is given by
$$k={F\over dU_{\rm el}/dF} =  {E^*\over q_r} {\bar F \over d\bar U_{\rm el}/d\bar F}$$
which gives
$$k =  \theta {E^*\over q_r} \left ({F q_r \over E^* h^0_{\rm rms} s^{1/2}}\right )^{1/(1+H)}\eqno(1)$$
where
$${1\over \theta } = {1+2H\over 1+H} \left ({3\over 4}\right )^{H/(1+H)} \left ({\beta^2 H\over 2-H}\right )^{{1/2(1+H)}} \pi^{(2-H)/(1+H)}(\kappa_0+\kappa_1 )={1\over \theta_0 } + {1\over \theta_1 }$$
where
$$\kappa_0 = {2\over 5}$$
$$\kappa_1 = \left ({2-H\over H}\right )^{1/2} {\gamma \over \beta \pi^2}$$
The stiffness per unit area $K=k/A_0 = k/L^2$:
$$K= \theta {E^* \over \pi \lambda_r} \left ({\lambda_r\over L}\right )^2  \left ({F q_r \over E^* h^0_{\rm rms} s^{1/2}}\right )^{1/(1+H)}$$

\begin{figure}[tbp]
\includegraphics[width=0.50\textwidth,angle=0]{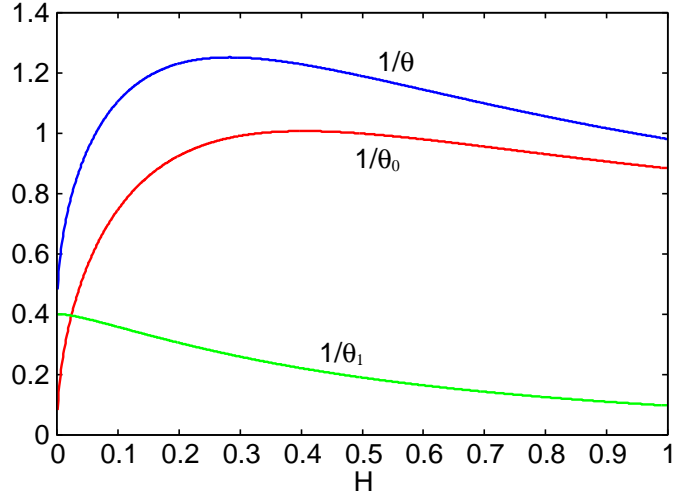}
\caption{The quantities $1/\theta_0$, $1/\theta_1$ and $1/\theta= 1/\theta_0+1/\theta_1$ are defined in the text.}
\label{1H.2invThete1.invTheta2.invSum}
\end{figure}

In Fig. \ref{1H.2invThete1.invTheta2.invSum}
we show $1/ \theta_0$, $1/ \theta_1$ and $1/\theta = 1/\theta_0+1/\theta_1$ as a function of the Hurst exponent $H$.
It is interesting to note that as $H\rightarrow 0$,
then $1/ \theta_0 \rightarrow 0$ while $1/ \theta_1$ remains finite, i.e., for the fractal dimension $D_{\rm f} = 3$
the stiffness is entirely determined by the  short-wavelength roughness in the macroasperity contact region.
Note also that since $q_r \approx \pi /L$, where $L$ is the linear size of the system,
the stiffness scales as $k\sim q_r^{-H/(1+H)} \sim L^{H/(1+H)}$ with the size of
the system. This is in contrast to the region where the $p\sim {\rm exp}(-u/u_0)$ relation holds
where the interfacial contact stiffness is independent of the
size $L$ of the system. Note also that the stiffness scales with the
rms roughness as $(h^0_{\rm rms})^{-1/(1+H)}$ while in the region where the $p\sim {\rm exp}(-u/u_0)$ relation holds
the stiffness is proportional to $(h^0_{\rm rms})^{-1}$.
For the Hurst exponent $H\approx 0.8$, which is typical in practical applications, $\theta \approx 1$, which appears to be in
good agreement with the prefactor found by Pohrt and Popov in their numerical simulation study~\cite{Pohrt12}.
The treatment
presented above can be generalized to obtain the distribution of stiffness values (at least approximately)
by calculating the distribution $P(R)$ of summit curvature radius $R$ (which is easy to do).

It is interesting to determine the critical force $F_{\rm c}$ such that for $F < F_{\rm c}$ one needs to use the finite
size region expression for the stiffness while for $F > F_{\rm c}$ the Persson expression is valid.
When the relation $p\sim {\rm exp} (-\bar u/u_0)$ is valid the stiffness
$$k={F \over u_0} = {F \over \gamma h^0_{\rm rms}}\eqno(2)$$
The critical force $F_{\rm c}$ is determined by the condition that $k$ given by (1) and (2) coincide. This gives
$$\theta {E^*\over q_r} \left ({F_{\rm c} q_r \over E^* h^0_{\rm rms} s^{1/2}}\right )^{1/(1+H)}={F_{\rm c} \over \gamma h^0_{\rm rms}}$$
or
$$F_{\rm c} = E^* {h^o_{\rm rms}\over q_r} s^{-1/2H} \left (\theta \gamma \right )^{(1+H)/H} \eqno(3)$$
Note that typically (for $H\approx 0.8$) $(\theta \gamma)^{(1+H)/H} \approx 0.2$. The prediction (3) for the switching between
the finite size region and the region where the stiffness is proportional to the loading force is in good agreement with our
simulation results. To show this let us first write
$${p_{\rm c} \over E^*} = {F_{\rm c}\over E^*L^2} =  {q_rh^0_{\rm rms} \over (q_r L)^2} s^{-1/2H}  \left (\theta \gamma \right )^{(1+H)/H}
= {q_rh^0_{\rm rms} \over \pi^2} \left ({\lambda_r \over L}\right )^2 s^{-1/2H}  \left (\theta \gamma \right )^{(1+H)/H}\eqno(4)$$
The surfaces we have studied in numerical simulations have the rms slope 0.1. To relate this to $q_rh^o_{\rm rms}$ which enters in (4) we use that
$$\langle (\nabla u )^2 \rangle = \int d^2q \ q^2 C(q) \approx  2\pi C_0 \int_{q_r}^{q_1} dq \ q^3 \left ({q\over q_r}\right )^{-2(1+H)}
\approx {Hs \over 1-H} (q_r h^o_{\rm rms})^2 \left ({q_1\over q_r}\right )^{2(1-H)}$$
or
$$q_r h^o_{\rm rms} \approx \langle (\nabla u )^2 \rangle^{1/2} \left ( {1-H\over Hs}\right )^{1/2} \left ({q_r\over q_1}\right )^{1-H}\eqno(5)$$
In the present case the rms slope is 0.1 and $q_0/q_1 = 1/4096$ and $H=0.7$ so that $q_r h^o_{\rm rms} \approx 5.5\times 10^{-3}$ for $q_r/q_0 =
L/\lambda_r =1$, and $q_r h^o_{\rm rms} \approx 1.3\times 10^{-2}$ for $q_r/q_0 = 8$.
Using this from (4) we get $p_{\rm c}/ E^* \approx 8\times 10^{-5}$ for $q_r/q_0 = 1$ and $p_{\rm c}/ E^* \approx 4\times 10^{-6}$ for $q_r/q_0 = 8$,
which is in good agreement with Fig. 1 in our paper.
For the surface with $H=0.3$ from (5) we obtain (for a surface with the rms slope 0.1)
$q_rh^o_{\rm rms}$ nearly $100$ times smaller than for $H=0.7$, which will shift
the cross-over force $F_{\rm c}$, between the two stiffness regions, with a similar factor to lower values, again in good agreement
with the numerical studies.
The results presented above differ from the conclusion of Pohrt and Popov who state that the power relation observed for small
applied forces is valid for all applied forces~\cite{Pohrt12,PRE}.
In particular, in Ref.~\cite{PRE} Pohrt et al state: ``It is the authors strong belief that the proportionality found
by Persson appears only in the first case described. Whenever
the surfaces are truly fractal with no cut-off wavelength, a
power law applies.'' The present study shows that this statement is incorrect and Fig. 1 in our letter clearly shows that the
contact stiffness cannot be described by a power law for all applied forces as this would correspond to a straight line on our
log-log scale.

As an example consider applications to syringes, where the relation between the squeezing pressure $p$ and the (average)
interfacial separation $\bar u$ (which determine the contact stiffness) is very important for the fluid leakage at the
rubber-stopper barrel interface. Consider the contact region between a rib of the rubber stopper and the barrel.
The width of the contact region (of order $w\approx 1 \ {\rm mm}$) defines the cut off wavevector $q_r = \pi /w \approx 3000 \ {\rm m}^{-1}$.
The Hurst exponent $H\approx 0.9$ and the rms roughness amplitude (including the roughness components with wavevector $q > q_r$)
is $h^o_{\rm rms} \approx 3 \ {\rm \mu m}$. The elastic modulus of the rubber stopper is typically $E\approx 3 \ {\rm MPa}$.
Using these parameters we get from (4): $p_{\rm c} \approx 1 \ {\rm kPa}$,
which is negligible compared to the pressure in the contact region between the rib of the rubber stopper and the barrel,
which is typically of order $\sim 1 \ {\rm MPa}$.

\section{Experiments}

The relation (1) as well as the above mentioned finite-size effect region has also been observed in experiments. In these
experiments a rectangular block of silicon rubber (a nearly perfect elastic material even at large strain) is squeezed
against hard, randomly rough surfaces. In this case no plastic deformation will occur, and the compression
of the rectangular rubber block, $(p/E')d$ (see below), which will
contribute to the displacement $s$ of the upper surface of the block, can be accurately taken into account. Such measurements were performed
in Ref. \cite{Lorenz}, and were found to be in good agreement with the theory (these tests involved no fitting parameters as the surface roughness
power spectrum, and the elastic properties of the rubber block, were obtained in separate experiments).
Here we show the result for the contact stiffness $K=-dp/d\bar u$ (not presented in Ref. \cite{Lorenz}) of one additional such measurement.

\begin{figure}[thbp]
\includegraphics[width=0.50\columnwidth]{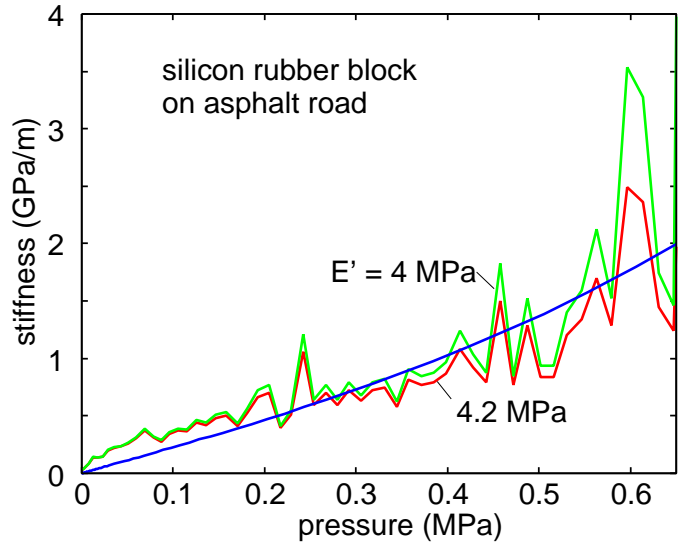}
\caption{\label{experiment}
The normal contact stiffness as a function of the applied nominal contact pressure for
a silicon rubber block (cylinder shape with diameter $D=3 \ {\rm cm}$ and height $d=1 \ {\rm cm}$)
squeezed against a road asphalt surface. The green and red lines are obtained from the measured
$p(s)$ relation using (7) with $E'=4.0 \ {\rm MPa}$ and $4.2 \ {\rm MPa}$ (see text) while the blue line is the
theory prediction.
}
\end{figure}

The experiment was performed for a silicon rubber block (cylinder shape with diameter $D=3 \ {\rm cm}$ and height $d=1 \ {\rm cm}$)
squeezed against a road asphalt surface with the rms roughness amplitude $0.63 \ {\rm mm}$ and the roll-off wavelength $\lambda_L \approx
0.3 \ {\rm cm}$ as inferred from the surface roughness power spectrum.
The squeeze-force is applied via a flat steel plate and no-slip of the rubber
could be observed against the steel surface or the asphalt surface.
We measured the displacement $s$ of the upper surface of the block as a function of the
applied normal load. Note that
$$s=(u_{\rm c}-\bar u)+(p/E')d\eqno(6)$$
where $E'$ is the effective Young's modulus taking into account the no-slip
boundary condition on the upper and lower surface, which was measured to be $E'=4.2 \ {\rm MPa}$ in a separate experiment where the
rubber block was squeezed between two flat steel surfaces. Using (6) gives
$$K=-{dp \over d\bar u} = -{dp \over ds} {ds\over d\bar u} = {dp \over ds} \left (1+{K d \over E'}\right )$$
or
$$K={K^* \over 1-K^* d/E'}\eqno(7)$$
where $K^* = dp/ds$. Using (7) in Fig. \ref{experiment}
shows the normal contact stiffness as a function of the applied nominal contact pressure
obtained from the measured $p(s)$ relation with $E'=4.2 \ {\rm MPa}$ (measured value)
and $E'=4 \ {\rm MPa}$ (to indicate the sensitivity of the result to $E'$). For very small contact pressures $K^* \approx 0$
so that the denominator in (7) is $\approx 1$ (and $K\approx K^*$ as assumed in Ref.
\cite{Pohrt12} without proof) and the result is insensitive to $E'$ as also seen in Fig. \ref{experiment}.
For large contact pressure the experimental data exhibits rather large noise (and great sensitivity to $E'$),
which originates from the increasing importance of the compression of the rubber block for large contact pressure. That is,
for large pressures the denominator in (7) almost vanishes, which implies that a small uncertainty in the measured $p(s)$ relation
(which determines $K^*$), or in $E'$,  will result in a large uncertainty in $K$ for large pressures.

The blue curve in Fig. \ref{experiment} is the theory prediction which is obtained
without any fitting parameter using the measured surface roughness power spectrum. For small contact pressure the contact stiffness
obtained from the measured data is larger than predicted by the theory,
but for nominal contact pressures typically involved in rubber applications (which are $\sim 0.4 \ {\rm MPa}$ as in tire applications,
or higher in most other applications) the finite size effects are not important.

\end{document}